\documentclass[a4paper,twocolumn]{article}
\pdfoutput=1

\usepackage[round]{natbib}
\usepackage{graphics}

\usepackage{mathtools}

\usepackage{fancyhdr}
\begin{document}
\fancyfoot[R]{\footnotesize To appear in Astrophysics and Space Science}
\fancyhead[L]{}
\fancyhead[R]{}

\title{Prudence in estimating coherence between planetary, solar and climate oscillations}


\author{Sverre Holm,\\
Department of Informatics, University of Oslo}
\maketitle

\begin{abstract}
There are claims that there is correlation between the speed of center of mass of the solar system and the global temperature anomaly. This is partly grounded in data analysis and partly in a priori expectations. The magnitude squared coherence function is the proper measure for testing such claims. It is not hard to produce high coherence estimates at periods around 15--22 and 50--60 years between these data sets. This is done in two independent ways, by wavelets and by a periodogram method. But does a coherence of high value mean that there is coherence of high significance? In order to investigate that, four different measures for significance are studied. Due to the periodic nature of the data, only Monte Carlo simulation based on a non-parametric random phase method is appropriate. 

None of the high values of coherence then turn out to be significant. Coupled with a lack of a physical mechanism that can connect these phenomena, the planetary hypothesis is therefore dismissed.
\end{abstract}



\section{Introduction}
\begin{figure}[tb]
\begin{center}
	\includegraphics[width=\columnwidth]{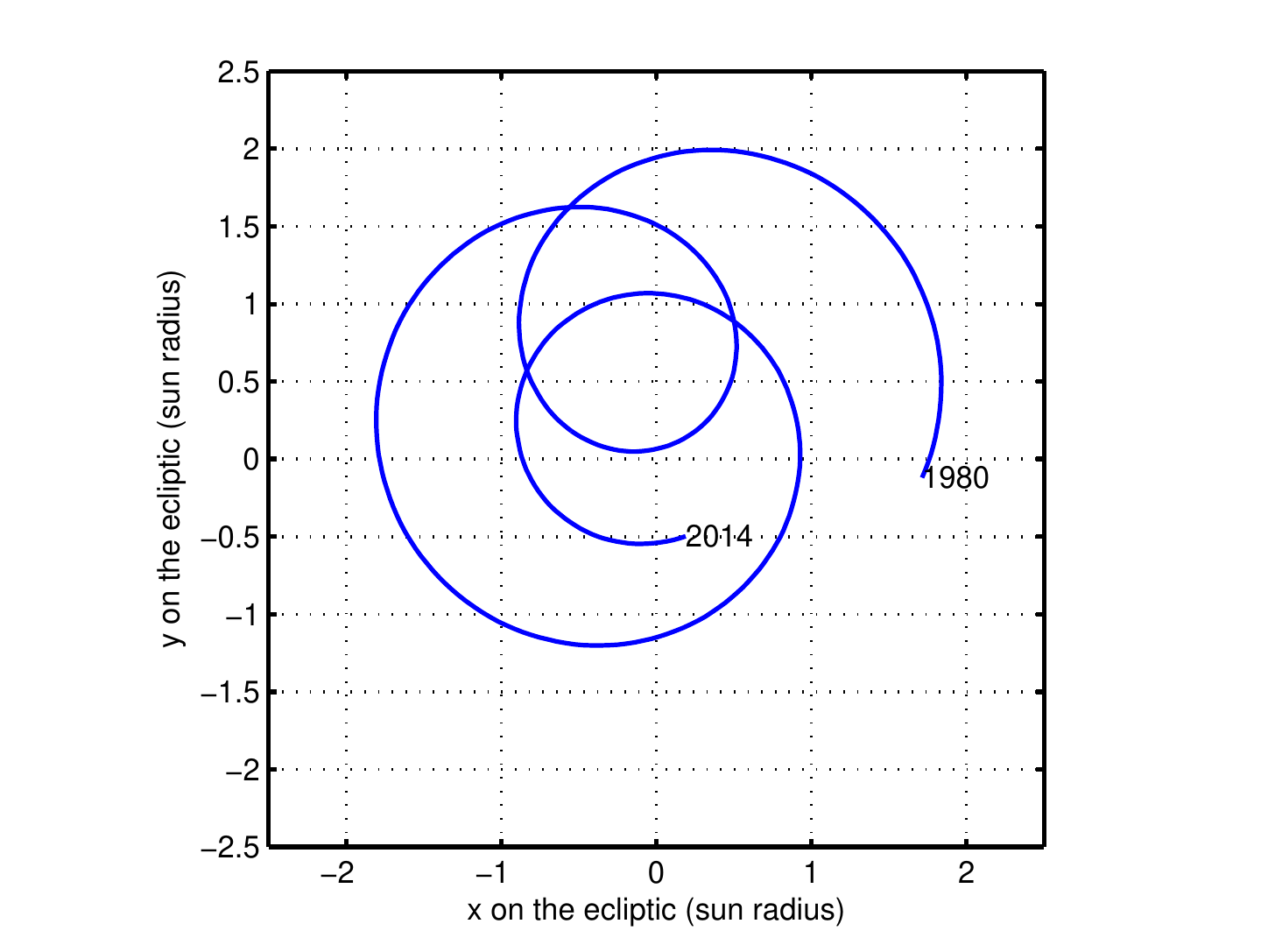}
	\caption{Orbit of the center of mass of the solar system from 1980 to 2014 generated by the Horizons system of JPL} 
	\label{fig:SolarOrbit}
\end{center}
\end{figure}

The center of the mass of the solar system moves up to about one sola diameter due to the influence of in particular the gas giants Jupiter and Saturn. A 2-D view of the orbit is shown in Fig.\ \ref{fig:SolarOrbit}. Can the signature of this orbit be found in the global temperature? 

In order to argue this case one needs  two kinds of evidence. The first and most important one is an explanation in terms of a physical model that explains why this orbit should influence processes in the sun which in turn will influence the temperature on earth. There is no such explanation which is generally accepted today, see e.g.\ \citet{callebaut2012influence}, despite the existence of alternative theories \citep{scafetta2012does} and recent discussions \citep{abreu2012there, cameron2013no, cauquoin2014no, poluianov2014critical, abreu2014response}. 

\begin{figure}[tb]
\begin{center}
	\includegraphics[width=.9\columnwidth]{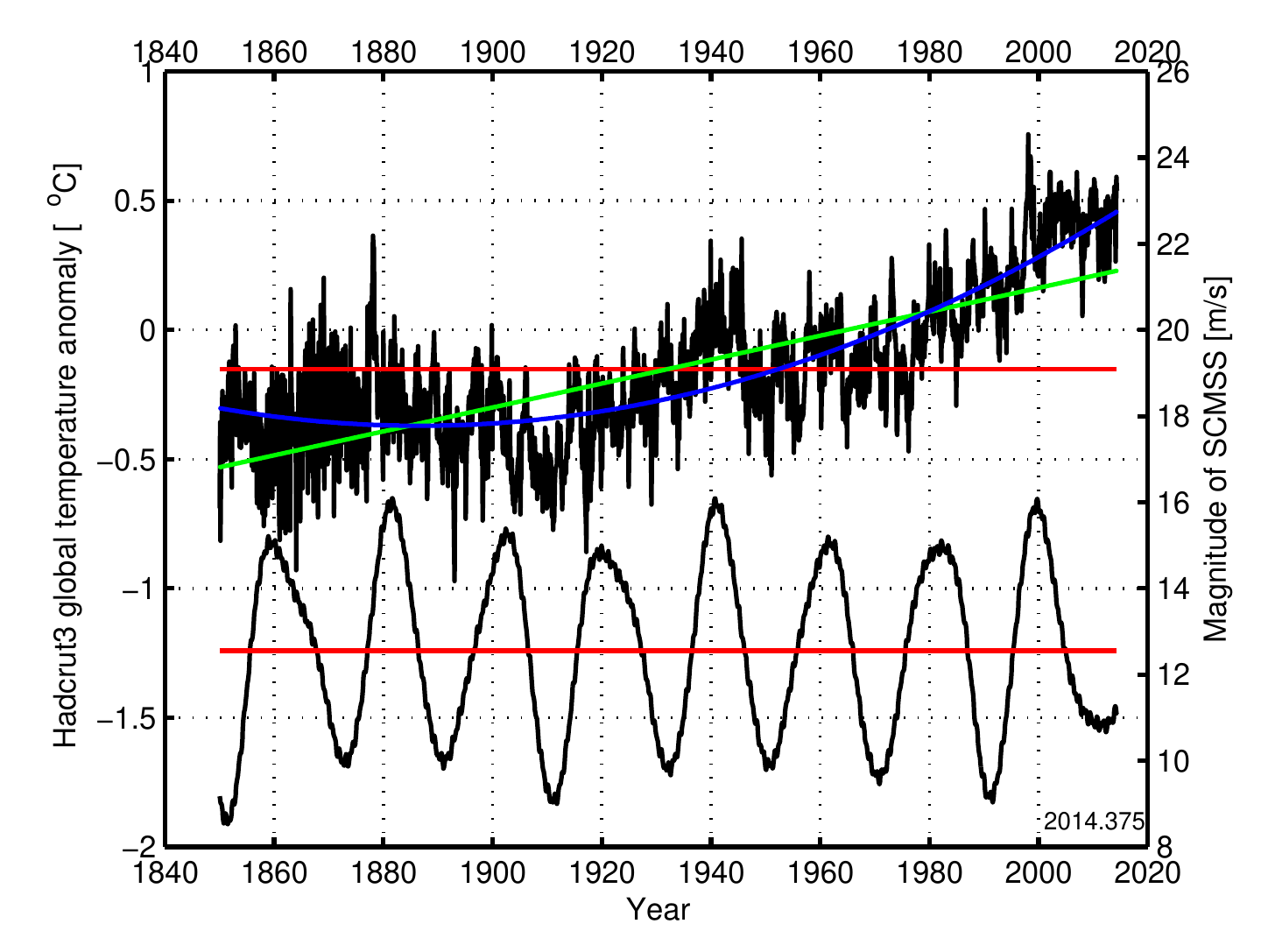}
	\caption{Global temperature anomaly HadCRUT3, \citet{brohan2006uncertainty}, upper curve, with mean, linear and quadratic fits and speed of center of mass of the solar system (SCMSS) generated by the Horizons system of JPL, lower curve, with mean value} 
	\label{fig:input}
\end{center}
\end{figure}

The second way is to compare data and look for correlation and coherence. This is the approach followed by \citet{Scafetta2010}. He compared the HadCRUT3 global temperature anomaly with the speed of the center of mass of the solar system (SCMSS). The raw data is plotted in Fig.\ \ref{fig:input} and runs from 1850 to May 2014. The most obvious feature of the temperature data (upper curve) is a gradual rise which is captured imperfectly by the linear and parabolic fits. The secondary feature is a pattern of oscillations which is what this paper is concerned with. Scafetta is to be commended for having drawn the attention to these oscillations and for raising various hypotheses for their origin. The most obvious feature of the SCMSS data in the lower part of Fig.\ \ref{fig:input} is a periodicity of about 20 years which is the synodic period of Jupiter and Saturn, i.e. the period for which their positions and that of the Sun are realigned. 

The main argument of \citet{Scafetta2010} was based on a comparison of power spectra which allegedly demonstrated coherence, and in particular at periods around 20 and 60 years. Then in \citet{holm2014alleged,holm2014corrigendum} I argued that this is only a qualitative comparison. There has for a long time been available a much better tool and this is the magnitude squared coherence (MSC) function. It is a frequency dependent normalized correlation value between 0 and 1 which is also sensitive to phase relationships. The MSC was therefore introduced to this problem and it was shown that in order to get a statistically reliable estimate of coherence from the 160 or so years of data available, one could not average over windows which were much longer than 40 years. This was due to the particular significance interval estimate that was used in that paper. It was therefore not possible reliably to resolve coherence at a period around 60 years. 

As a response to this, it was argued in \citet{scafetta2014DiscussionCritiques} along three lines. First the similarity of spectral estimates of climate and planetary series was again used as an argument as Fig.\ 6B from \citet{Scafetta2010} was reproduced as Fig.\ 12 of \citet{scafetta2014DiscussionCritiques}. He also reproduced the results from an undocumented coherency test. Contrary to him, I actually see the opposite of coherence in this figure as the spectral lines do not overlap very convincingly at frequencies around the 20 and 60 year periods. 

In the rest of Scafetta's paper the MSC of  \citet{holm2014alleged} has now been adopted rather than a comparison of spectral estimates. His second argument was that one needs to improve the MSC estimator based on averaged windowed periodograms and cross-spectra of \citet{holm2014alleged} and replace it with the minimum variance distortion-less (MVDR) MSC estimator \citep{benesty2006estimation}. Third he argued that with this high resolution MSC estimator, one could increase the window length to 110 years. With windows of this length, a line around 60 years should be resolvable, and such a line came out of his analysis as well.

I don't see this as a discussion of right or wrong  methodology, at least not as long as one is talking about estimating the MSC. It is rather about significance and questions such as if a high coherence value automatically implies that it has high significance, and how important the reliability of the coherence estimator's amplitude is. Since there is a trade-off between spectral resolution and confidence in the amplitude estimate, given the limited time-span, then indirectly this is also a question of resolution. My main aim in this paper is to show that the primary difference between \citet{scafetta2014DiscussionCritiques} and \citet{holm2014alleged} is the weight placed on these properties.

A secondary difference is the emphasis placed on \emph{a priori} expectations. They are fine for framing a hypothesis, but not necessarily for defending it when data goes against the hypothesis. But this is what \citet{scafetta2014DiscussionCritiques} does when he invokes Kepler's first and second laws as well as Kepler's \emph{Mysterium Cosmographicum} (1596, 1621) and \emph{Harmonices Mundi} (1619)  to argue against my analysis since it didn't find 60 year periodicities of the strength that he anticipated. 

Further, it is not at all obvious that Kepler's contributions, beyond his three laws, are so useful. Here is what a critical historian of science has written: \emph{"The three major gems in his works on astronomy lay in a vast field of errors, of irrelevant data and details, of mystical fantasies, of useless speculations, of morbid detours of self analysis, and, last but not least, of an organismic and animistic conception of the world and its processes"} \citep{jaki1977planets}. Anyway, to me this is a simple question of analysis of the data and not of any \emph{a priori} expectations. The SCMSS data used in \citet{Scafetta2010} was clearly not available to Kepler and contains whatever it contains regardless of what expectations one may get from reading Kepler. Could it also be that the seemingly arbitrarily chosen scalar SCMSS may be a poor choice if the objective is to illustrate the 60 year periodicity of the planetary system, and that for instance it is easier to find it if the 3D vector information is used as well?

In this paper I will take it as a premise that comparisons of spectral peaks cannot be anything but a qualitative indicator of coherence and therefore argue only based on the magnitude squared coherence. I then discuss the effect of detrending on the global temperature data. A standard wavelet coherence estimation routine which is widely used in climate and geophysical data analysis is then applied to the data. I also demonstrate that it is not necessary to introduce a new MVDR MSC estimator to see peaks around 60 years, the periodogram based estimator may also show that if the window length is increased. In order to validate these peaks their statistical significance must be found. I discuss four ways of doing that and end up with a random phase method which has been designed to be applicable to serially correlated data as one has here. This is a new method compared to that used in \citet{holm2014alleged}. I will demonstrate that with common values of significance levels it is hard to argue for the statistical significance of coherence between the data sets for both the 20 year and the 60 year periodicities.
\begin{figure}[tb]
\begin{center}
	\includegraphics[width=.9\columnwidth]{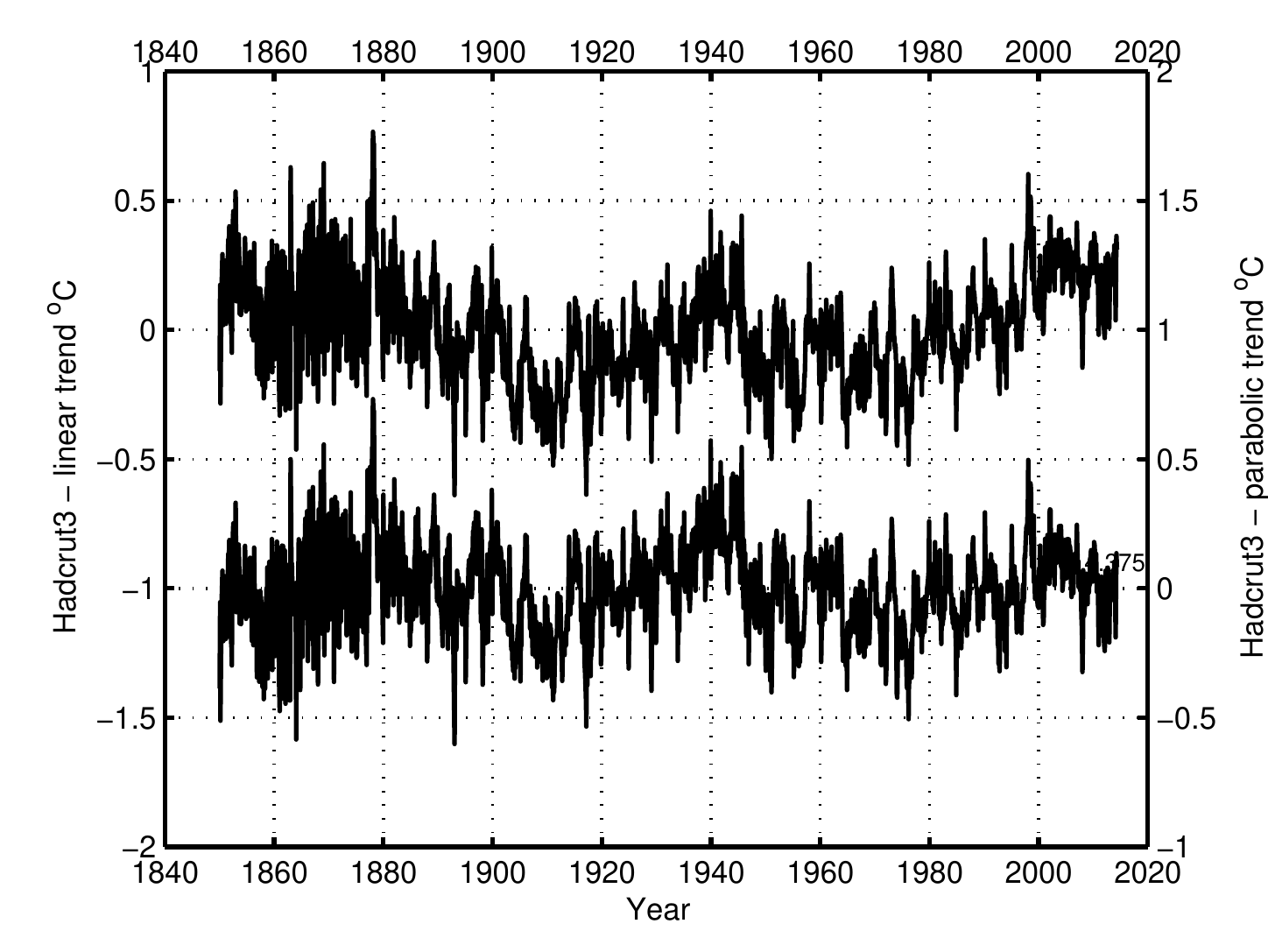}
	\caption{Detrended global temperature anomaly. Upper: linear detrending as used in this study, lower: parabolic detrending} 
	\label{fig:detrended}
\end{center}
\end{figure}
%
\section{Estimation of Magnitude Squared Coherence}

Two independent estimation methods, wavelet coherence and periodogram based coherence, are used here. An alternative which is not discussed is multitaper MSC estimation. It has been used for instance to demonstrate coherence between the global temperature and CO2 \citep{kuo1990coherence}. Another possibility is the 
MVDR estimator for MSC of \citet{benesty2006estimation} used by \citet{scafetta2014DiscussionCritiques}. These results are straight-forward to reproduce and the method was used by Scafetta because it was believed that the periodogram based MSC estimator could not resolve the interesting line at a 60 year period.  As will be shown here this is not the case, so therefore the relatively uncharacterized MVDR estimator which also depends on the setting of a regularization parameter is not used here. In general statistical properties are much harder to find for MVDR-based estimators. This is the case also in other fields where we have experience with MVDR or it equivalents, the Capon method and adaptive spectral estimation \citep{synnevag2007adaptive}.

\subsection{Detrending}

The MSC results are sensitive to the kind of detrending applied to the global temperature data. Alternatives are to subtract the mean, to subtract a linear trend, or to subtract a parabola. These curves fitted to the raw data are shown in Fig.\ \ref{fig:input}. Subtraction of the mean is obvious to do in any spectral method in order to prevent strong components at 0 Hz to leak into the low frequency parts. The most critical period in question is at around 60 years. This is a frequency of only $1/60$ year$^{-1}$ which is very close to 0 and in the frequency range which is affected most by detrending. 

The safest way to justify linear and parabolic detrending is to argue in terms of a physical model for the data. Lacking that, another criterion could be to test for sensitivity. If a result is critically dependent on one particular detrending, then the result would be dubious. The temperature data detrended by a linear fit and a parabolic fit were therefore plotted in Fig.\ \ref{fig:detrended}. Compared to the raw data in the upper panel of Fig.\ \ref{fig:input} it appears that in particular the lower series, detrended by a parabolic function, shows a tendency to amplify a low frequency oscillation in particular in the range 1850--1890 but also after 2000. 

These differences can be seen in the  MSC estimates also. Parabolic detrending enhances low frequency periodicites more than the other methods whether the periodogram based method or the MVDR method is used for MSC estimation. This will be commented on later also. In my view, parabolic detrending adds uncertainty to the result. Are we seeing phenomena that are in the data or are we seeing the result of an interaction between the detrending and the data?

A middle way, using subtraction of the mean for the SCMSS data and linear detrending for the temperature data has therefore been chosen here. The latter is also justified by the argument in \citet{ebisuzaki1997method} that linear detrending removes the influence of potential unresolved low frequencies. The careful analysis of linear detrending on temperature data in \citet{kuo1990coherence} also justifies it.

\subsection{Statistical significance}
An analytical expression for the statistics of the MSC estimator exists for the case of an integer number of non-overlapping segments and Gaussian  uncorrelated (white) data. To test the null hypothesis of no coherence, the true value can be set to zero. In that case an independence threshold can be found \citep{carter1987coherence, wang2004optimising}. The method was used in \citet{holm2014alleged}. Since in particular the SCMSS data is highly correlated due to its periodic nature,  this will tend to indicate coherence where in reality there is none.

No analytical expression for the statistics exists in case of correlated data so one has to resort to Monte Carlo simulation to find a test for the null hypothesis. In order to generate a data set to test against, there are several possibilities. One way is to fit an AR(1) model to each data set and generate simulated data sets from this red noise model \citep{grinsted2004application}. This method will also indicate coherence when there is none when there are serially correlated and in particular periodic signals present \citep{poluianov2014critical}. 

The non-parametric random phase method of \citet{ebisuzaki1997method} is better as it was developed for serially correlated data. Several studies have used this method \citep{moron2008weather, traversi2012nitrate, poluianov2014critical}. It consists of Fourier transforming a data set, randomizing the Fourier phases while maintaining the complex conjugate symmetry of the frequency domain data so that the inverse transform is still real, and doing an inverse transform to get back to the time domain. \citet{ebisuzaki1997method} recommends that the pair-wise coherence between a large set (1000 in our case) random phase versions of time series 1 and the original time series 2 is found, and then that the results are sorted and the 95\% percentile is plotted along with the coherence of the two datasets. That paper also showed that as an AR(2)-process became more and more periodic, this method tends to assign coherence where there is none. Therefore we have adopted the method of the Appendix of \citet{traversi2012nitrate} where for each iteration also the coherence between a random phase version of series 2 and the original series 1 is found and the maximum of the two coherence functions is found for each step. The use of an additional phase criterion, as proposed there, is not used here.

The Monte Carlo simulation in \citet{scafetta2014DiscussionCritiques} is yet another way to attack this problem. However, since it is based on comparing the SCMSS with simulated sinusoids in additive independent noise at the expected frequencies, it is based on something different from  a null hypothesis. It starts with the assumption of coherence and tests its significance rather than starting with the hypothesis that there is no coherence. This goes against the principle of prudence in my opinion. It also depends on the value of a signal to noise ratio parameter. The setting of this parameter is not discussed at all in \citet{scafetta2014DiscussionCritiques} and an arbitrary value is just used without justification. Finally, this additive noise approach, which some might say is rather naive, is not even considered in papers that discuss how to find confidence intervals for serially correlated data \citep{ebisuzaki1997method}. 
\subsection{Wavelet coherence analysis}
\begin{figure}[t]
\begin{center}
	\includegraphics[width=0.9\columnwidth]{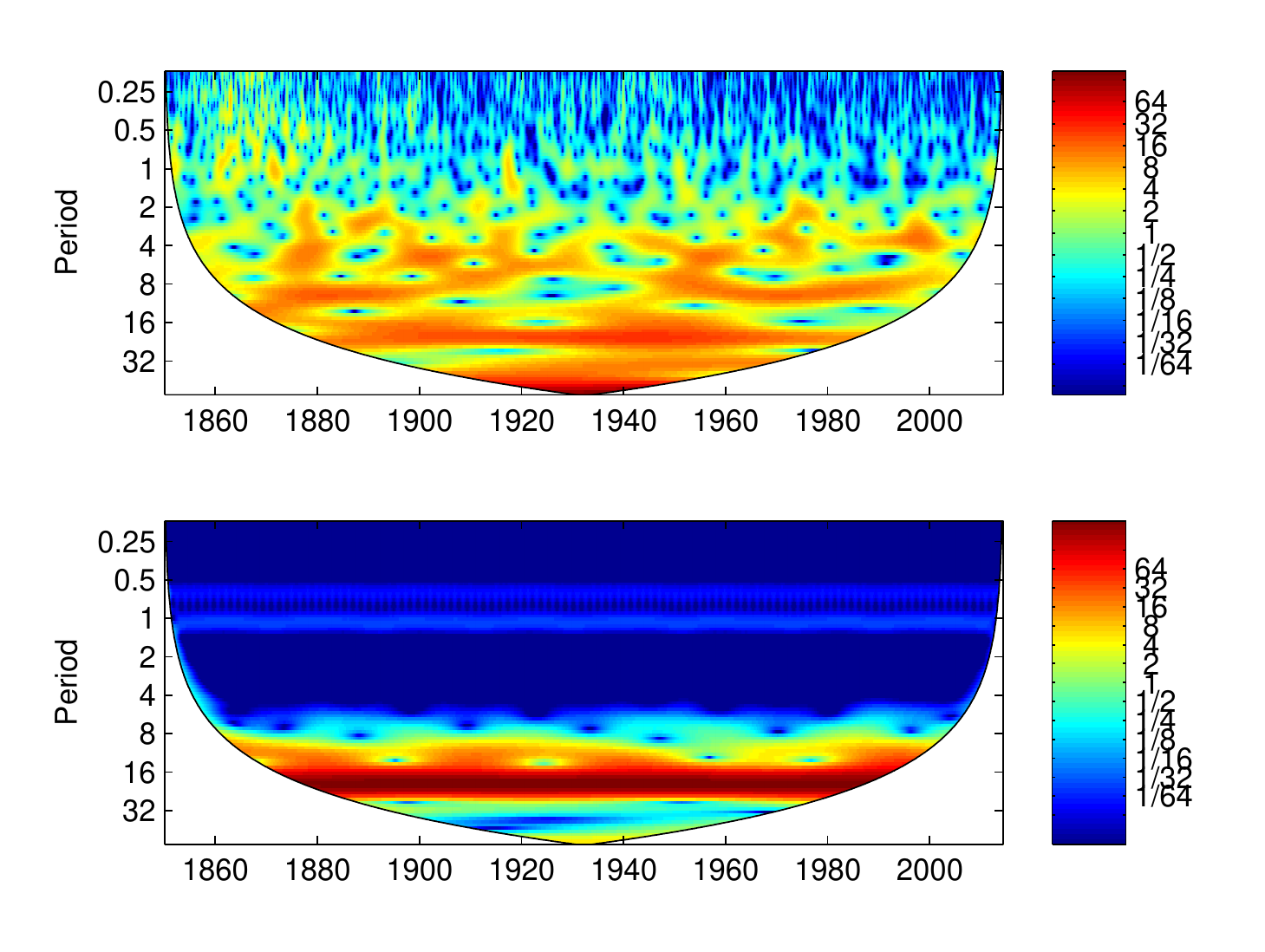}
	\caption{Wavelet spectra of linearly detrended HadCRUT3 (top) and zero-mean SCMSS (bottom)}
	\label{fig:Wavelet}
\end{center}
\end{figure}
\begin{figure}[tb]
\begin{center}
	\includegraphics[width=0.9\columnwidth]{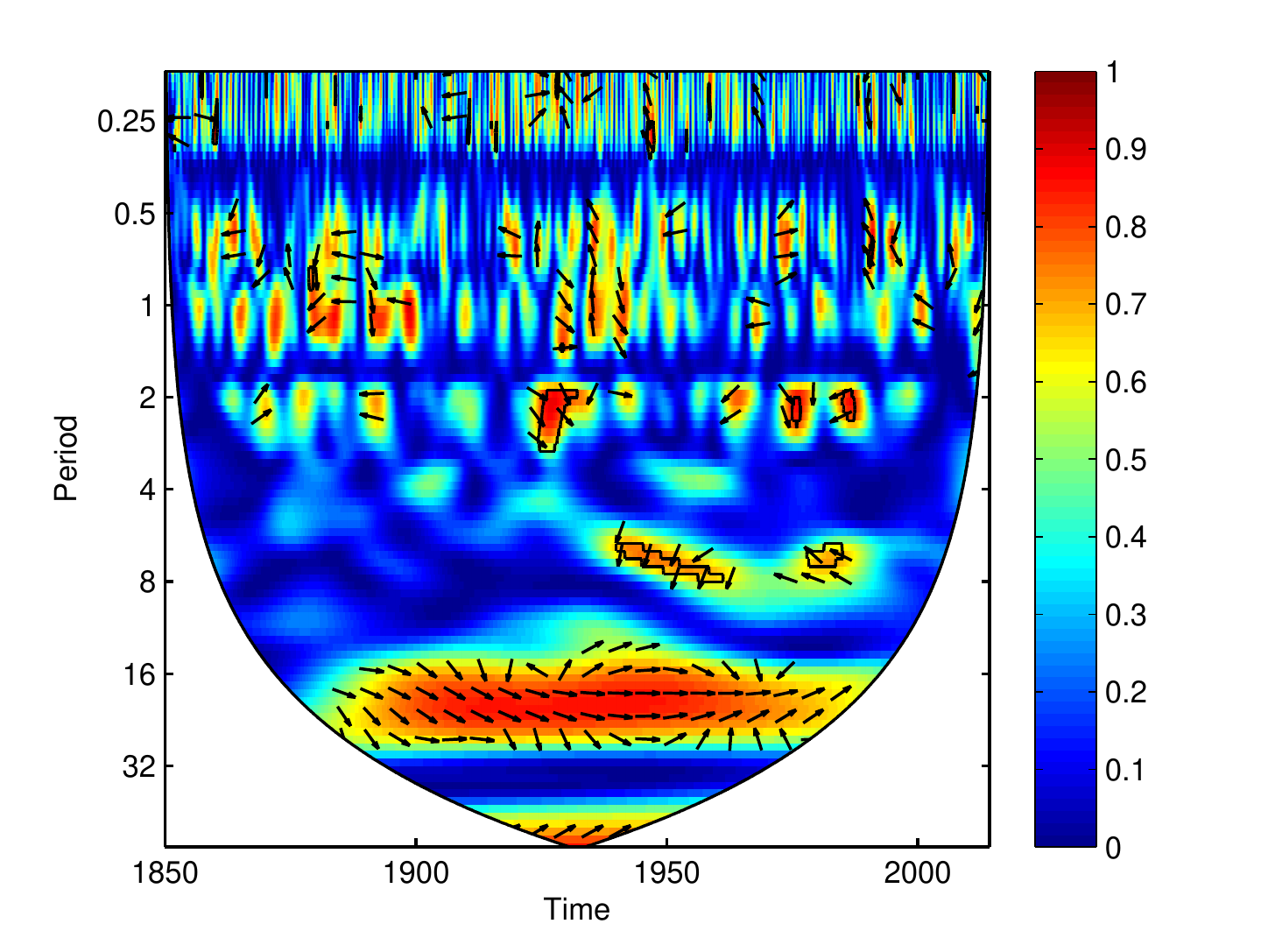}
	\caption{Wavelet magnitude squared coherence estimate for  linearly detrended HadCRUT3 vs zero-mean SCMSS. The contour shows the 95\% significance level based on the random phase model of \citet{traversi2012nitrate, poluianov2014critical}. The arrows show relative phase relationships with an arrow pointing to the right indicating in-phase variation}
	\label{fig:WTC}
\end{center}
\end{figure}
The wavelet estimation software of  \citet{grinsted2004application} has been used for many different climate related and astrophysical applications. It estimates spectra and magnitude squared coherence using the Morlet wavelet. In order to show where  the time-frequency space is free of edge artifacts, the data is only plotted inside the cone of influence, highlighting the lack of data for estimating components at high periods.

This method is applied to the detrended data of Fig.\ \ref{fig:input}. It can be seen that the wavelet spectra of Fig.\ \ref{fig:Wavelet} are in general agreement with the spectral analysis of these data sets in \citet{holm2014alleged,holm2014corrigendum}. A line around 60 years is very visible in the HadCRUT3 spectrum, but barely visible in the SCMSS spectrum. In addition the SCMSS spectrum has a strong component around 20 years which is also visible in the HadCRUT3 spectrum. 

The wavelet squared coherence function is shown in Fig.\ \ref{fig:WTC}. It is based on averaging of  two independent values in the time dimension and in the scale dimension, so the number of independent averages is  $N_d=4$ (see Appendix of  \citet{torrence1999interdecadal}). The confidence interval based on the random phase method from the Appendix of \citet{traversi2012nitrate} and \citet{poluianov2014critical} is used. There are only a few areas with significant coherence, like period $\sim 8$ years around 1950.  There is quite large coherence in the 17-22 and the 50-60 year ranges, but it is not large enough to reach above the 95\% significance level. This is different from the result of \citet{holm2014alleged} where coherence in the range 15-–17 years was found to be just above the significance threshold. The lack of significant coherence in Fig.\ \ref{fig:WTC} is due to the use of a significance interval estimate which is better suited for the serially correlated data.
%
%
%
%
\subsection{Periodogram based coherence estimate}
\begin{figure}[tb]
\begin{center}
	\includegraphics[width=.9\columnwidth]{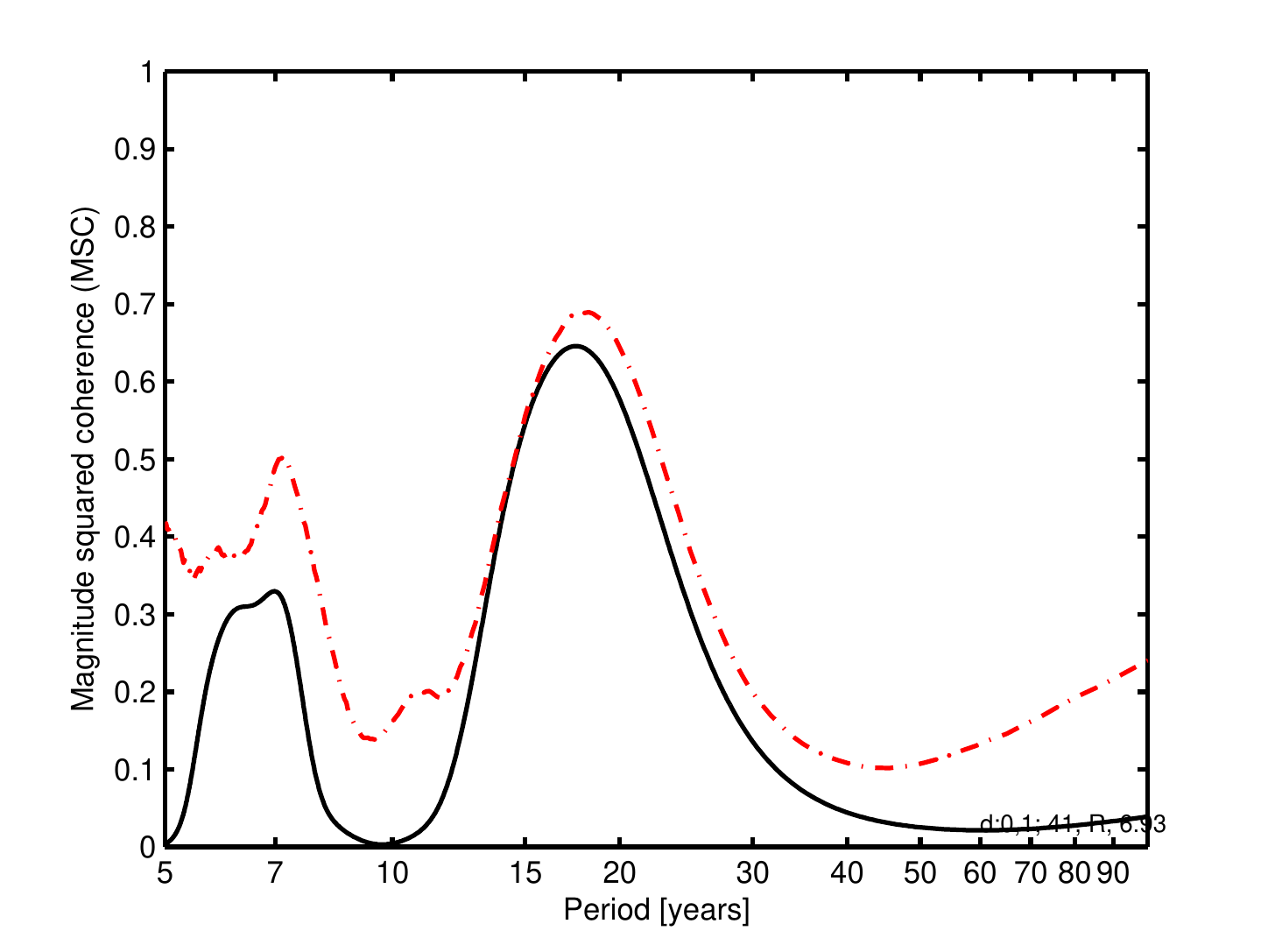}
	\caption{Periodogram  magnitude squared coherence of linearly detrended HadCRUT3 and zero-mean SCMSS found by averaging 13 overlapping segments of length 41 years, $N_d \approx 6.4$ independent averages under a white noise assumption. The 95\% significance threshold is shown as a dash-dotted line}
	\label{fig:MSC41}
\end{center}
\end{figure}
\begin{figure}[tb]
\begin{center}
	\includegraphics[width=.9\columnwidth]{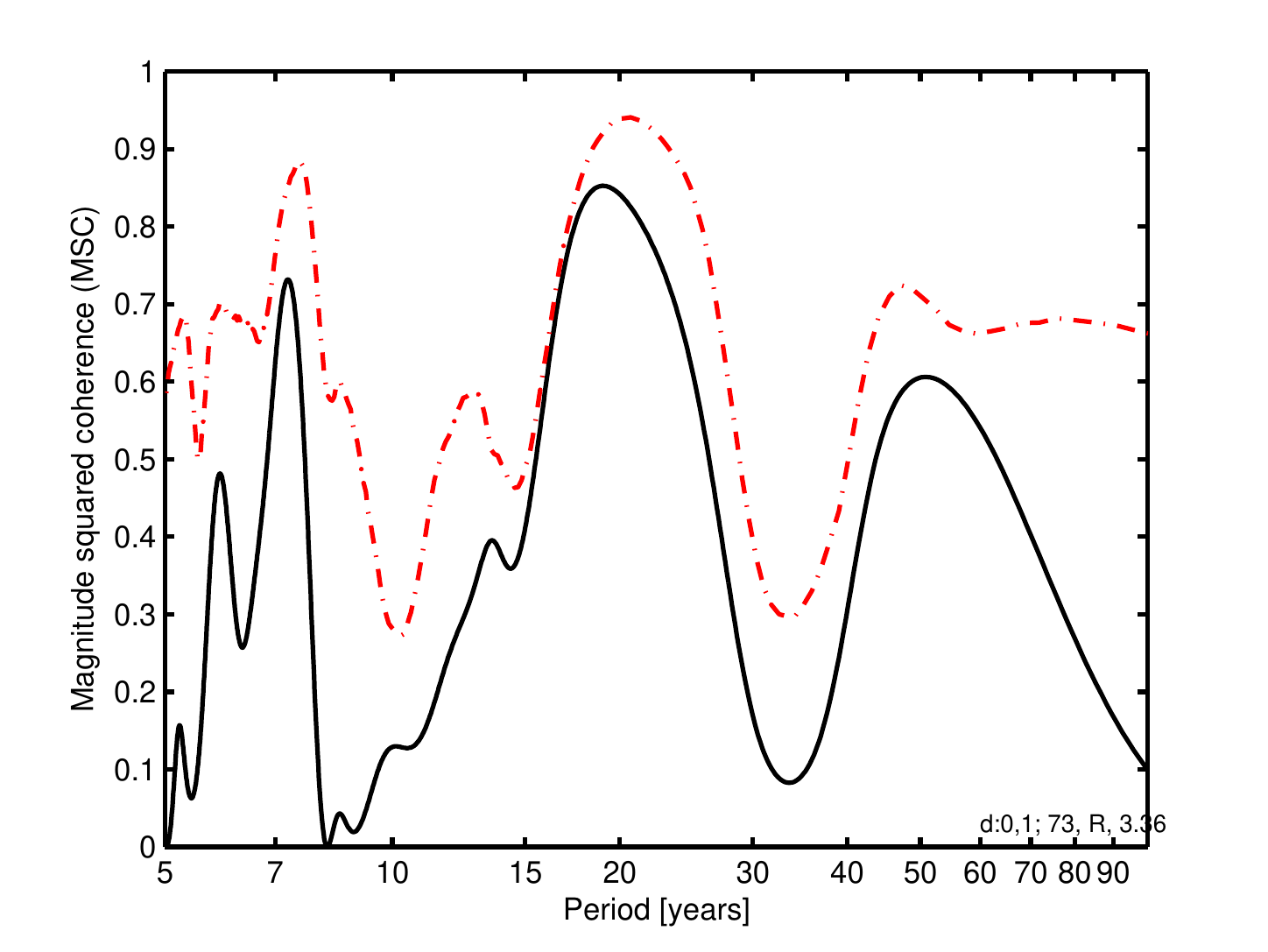}
	\caption{Periodogram magnitude squared coherence of linearly detrended HadCRUT3 and zero-mean SCMSS found by averaging 6 overlapping segment of length 73 years, $N_d\approx 3.4$ independent averages under a white noise assumption. The 95\% significance threshold is shown as a dash-dotted line}
	\label{fig:MSC73}
\end{center}
\end{figure}
\begin{figure}[bt]
\begin{center}
	\includegraphics[width=.9\columnwidth]{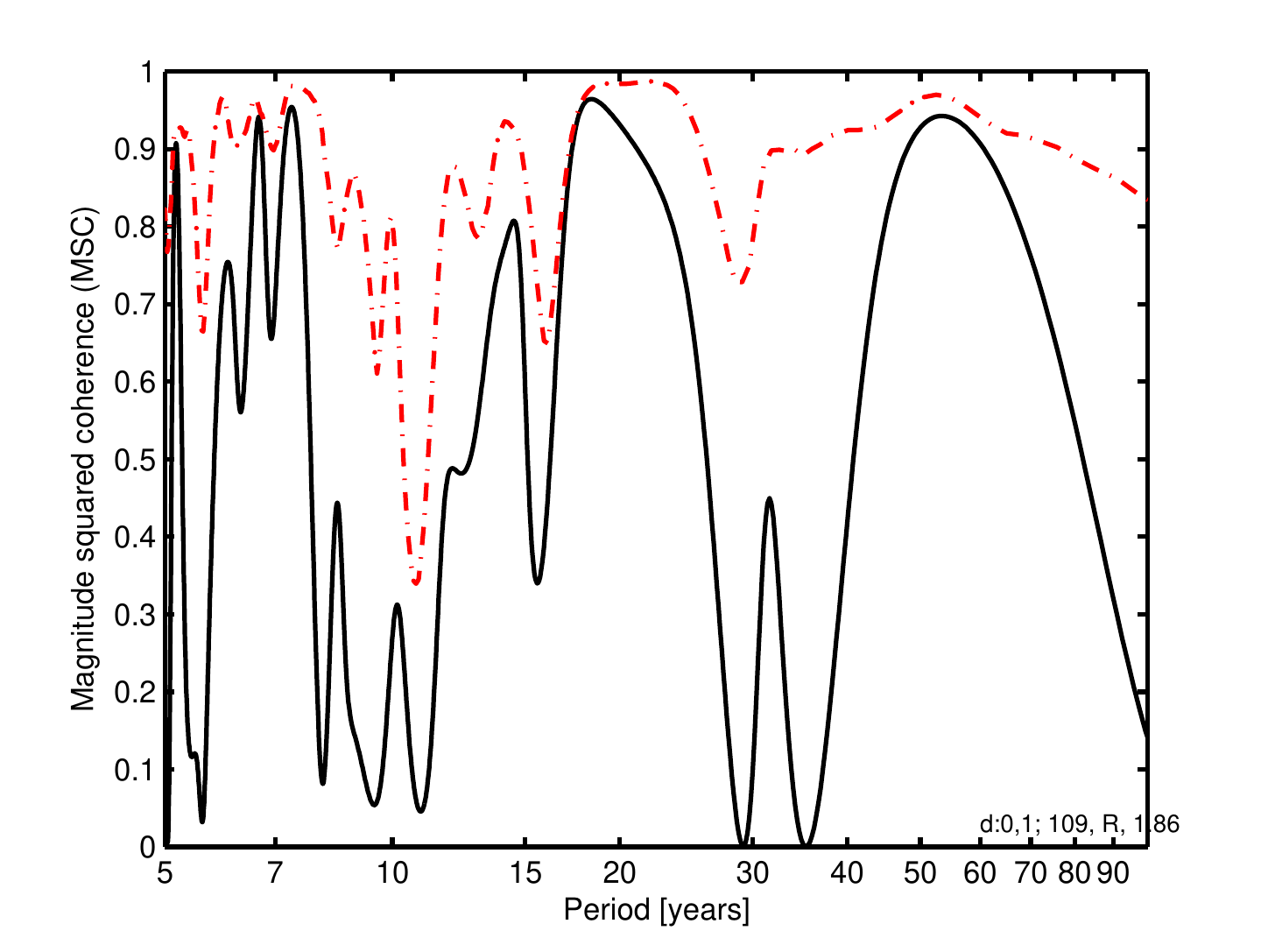}
	\caption{Periodogram  magnitude squared coherence of linearly detrended HadCRUT3 and zero-mean SCMSS found by averaging 3 overlapping segments of length 109 years, $N_d\approx 1.9$ independent averages under a white noise assumption.  The 95\% significance threshold is shown as a dash-dotted line}
	\label{fig:MSC109}
\end{center}
\end{figure}
In the periodogram based method for estimating magnitude squared coherence, the detrended time series is divided into short equal-length windowed segments. Then power and cross spectra are found by averaging over all the segments. The MSC estimate is found as the ratio of the squared cross spectrum estimate and the power spectra. The number of averages is given by the number of windowed segments. In order to utilize the data well, considerable overlap between segments is used. In our case we use a Kaiser window with $\beta=6$ and the overlap is $OL=75 \%$. That means that the number of independent averages, $N_d$, which determines confidence intervals, is considerably lower than the number of averages. See the Appendix for details of the algorithm and its properties.

In \citet{holm2014alleged} the MSC was plotted for window lengths 20 and 30 years, and results were given in table format for length 40 years also. Here I don't preselect the window length but rather find it by increasing it until the required number of segments fills all the available data  as much as possible giving window lengths like 41, 73, and 109 years. Despite this, some of the data will have to be discarded, but this is done from the beginning rather than from the end of the dataset. The reasoning is that the data from 1850 is both  less reliable and also less interesting than recent data. See Fig.\ \ref{fig:shifts} in the Appendix for an illustration of this.

The magnitude squared coherence using a window length of 41 years is shown in Fig.\ \ref{fig:MSC41}. The coherence in the 15--17 year range almost touches the 95\% confidence level, but is below it. One would have expected the solid MSC curve to reach well above the dash-dotted 95\% significance level if there was significant coherence.There is of course no disagreement between \citet{scafetta2014DiscussionCritiques} and I that windows of length 30--40 years are too short for finding coherence between time series on the 60 year scale. But as segment lengths are increased from 41 to 109 years, the number of independent averages, $N_d$, goes from 6.4 to 1.9. Due to the correlated nature of the data, this is the upper limit of $N_d$, and the actual number of independent averages is even lower. This number is so low  that significance levels based on the analytical model of \citet{holm2014alleged}  were not considered to be reliable. Therefore the window length was not increased beyond that in the first study.

Now that significance levels are found by Monte Carlo simulation, larger segment lengths like 73 and 109 years can give meaningful results. This is shown in Figs.\ \ref{fig:MSC73} and \ref{fig:MSC109}. It is seen that a peak in the 50--60 year range starts to appear in addition to the one around 19 years. These results are similar to Fig.\ 10 with 110 years segment length of \citet{scafetta2014DiscussionCritiques} with respect to the peak around 60 years, but not identical to it. 
The coherence estimate here is below the 95\% significance level, in particular in the 50--60 year range. 

The MSC peak at about 60 years in Fig.\ \ref{fig:MSC109} is above 0.9 while Scafetta's was below 0.8. The parabolic detrending of his Fig.\ 10 has actually enhanced the peak, and a change to linear detrending, like we use, will diminish Scafetta's MVDR peak even further to around 0.7.  Despite the higher value of our peak compared to \citet{scafetta2014DiscussionCritiques} it turns out not to reach above the dash-dotted 95\% significance level.

The period of the main peak in Figs.\ \ref{fig:MSC41} - \ref{fig:MSC109} is 17.5, 19, and 18.4 respectively. This is slightly less than 20 years, the peak value of the SCMSS spectrum. However, due to the ratio in the definition, Eq.\ \ref{Eq:msc}, the actual peak also depends on the cross spectrum as well as the resolution of the spectral estimates. Also it is evident from the upper panel of Fig.\ \ref{fig:Wavelet} that the peak position in the 20 year range for the temperature data varies with time, and this variation is smeared out in the MSC estimate. 
 
\section{Discussion and conclusion}
It is not hard to find high peaks in the magnitude squared coherence estimate in the 15-20 year range as well as in the 50-60 year range when the speed of center of mass of the solar system is compared to the global temperature anomaly. This is the case for the two independent methods, the wavelet coherence estimator of Fig.\ \ref{fig:WTC} and the periodogram based estimator of Figs.\ \ref{fig:MSC41} - \ref{fig:MSC109}. At first glance this seems to be in agreement with \cite{scafetta2014DiscussionCritiques}.

However, an estimate of coherence of high value does not mean that there is a coherence of high significance. Therefore a central point of this paper has been a discussion of significance levels. The challenge is that the data here is periodic. The four methods for assessing significance were:
\begin{enumerate}
\item Null hypothesis testing via Monte Carlo simulation based on the non-parametric random phase method for serially correlated data \citep{ebisuzaki1997method, traversi2012nitrate}
\item Null hypothesis testing based on an analytical expression for independence level derived for white Gaussian data \citep{holm2014alleged}
\item Null hypothesis testing via Monte Carlo simulation based on a red noise model \citep{grinsted2004application}
\item Testing via Monte Carlo simulation based on a simple model with sinusoids in additive noise
with an unknown signal to noise that needs to be set 
\citep{scafetta2014DiscussionCritiques}
\end{enumerate}
These methods will give widely different results with respect to significance. Prudent use of methods implies that conclusions only can be drawn from the first model as it is the only one that fits the data, as has been argued here. That means that coherence around 20 year and 60 year periods cannot be considered to be trustworthy. This strengthens the conclusion of \citet{holm2014alleged}.


The principle of prudence in addition to the lack of an accepted physical mechanism, then dictates the conclusion that one cannot say that there is any coupling between the movement of the center of the solar system and the global temperature data. It is more credible to look for down to earth explanations 
in e.g.\ oscillations in the atmospheric general circulation model, the Pacific decadal oscillation, or the Atlantic multi-decadal oscillation due to the meridional overturning circulation \citep{Gao2014landSurfaceTemperature, Sun2015delayedoscillator, 2015YaoHiatus}.
%
%

\section*{Acknowledgement}

Wavelet coherence software was provided by Aslak Grinsted and Stepan Poluianov. The software for randomization of phase was written by Vincent Moron. Thanks to Fritz Albregtsen, Knut Liest{\o}l, and Bj{\o}rn Samset for valuable comments.

\appendix
\section{Estimation of magnitude squared coherence}

The magnitude squared coherence (MSC) is defined as a normalized cross spectrum:
\begin{equation}
C_{xy}(f) = \frac{|P_{xy}(f)|^2}{P_{xx}(f)\cdot P_{yy}(f)},
\label{Eq:msc}
\end{equation}
where $P_{xx}(f)$ and $P_{yy}(f)$ are the power spectra of the two time series and $P_{xy}(f)$ is the complex cross spectrum. $C_{xy}(f)$ will always be between 0 and 1. For the estimate both time series are divided into $N$ segments which are Fourier transformed and averaged:
\begin{equation}
\hat{C}_{xy}(f) = \frac{|\sum_{n=0}^{N-1}X_n(f)Y^*_n(f)|^2} {\sum_{n=0}^{N-1}|X_n(f)|^2 \sum_{n=0}^{N-1}|Y_n(f)|^2},
\label{Eq:msc2}
\end{equation}
A case that demonstrates the need for averaging is to let $N=1$. Then the coherence estimator of Eq. \ref{Eq:msc2} will always output unity no matter what the input signals are and therefore be severly biased upwards. Such upward bias is typical when the number of averages is low.

\begin{figure}[tb]
\begin{center}
	\includegraphics[width=.9\columnwidth]{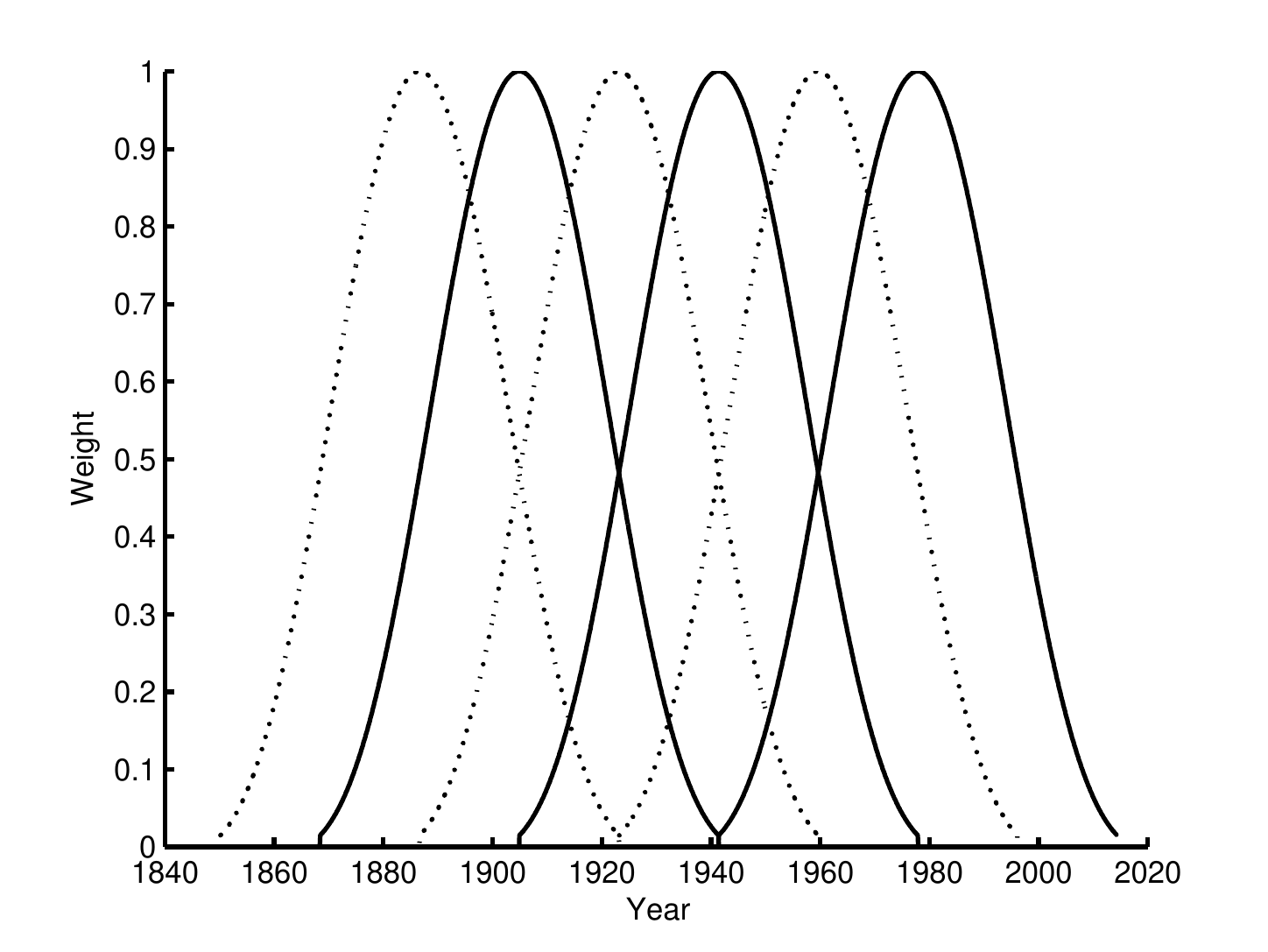}
	\caption{$N=6$ shifted and weighted windows with 75\% overlap, segment length $N_{seg}=73$ years}
	\label{fig:shifts}
\end{center}
\end{figure}

For averaging, each segment in time is weighted before Fourier transformation. Each segment overlaps with its neighbors as depicted in Fig.\ \ref{fig:shifts} for a Kaiser window with $\beta=6$.
Since the window  tapers the ends heavily down, $OL=75$\% overlap is used between the $N$ segments. The number of segments is:
\begin{equation}
N = 1 + \lfloor{\frac{N_{tot}-N_{seg}}{N_{seg}(1-OL)}}\rfloor
\label{Eq:N}
\end{equation}
where $N_{tot}$ is the total number of  samples and $N_{seg}$ is the segment length.

The  actual value for the number of independent averages, $N_d$, will be less than $N$ when there is overlap, but  statistics for the MSc estimator is only known for the case of an integer number $N_d\ge 2$ of non-overlapping segments \citep{carter1987coherence, wang2004optimising}.  However, one can assume that the reduction in the number of independent averages is similar to that which happens when estimating a spectrum \citep{holm2014alleged}. In the case of an overlap of $OL=75\%$, the number of independent averages, $N_d$, is estimated from the correlation of  the data window at the $p=25$, 50, and 75\% points, $c(p)$, \citet{harris1978use}, assuming that the input data is white:
\begin{align}
\begin{split}
N_d^{-1} &= \frac{1}{N} [1+2 c^2(0.75) + 2c^2(0.5)]\\
&-\frac{2}{N^2}[c^2(0.75) + 2c^2(0.5) + 3c^2(0.25)]
\end{split}
\label{Eq:Nd}
\end{align}
The ratio  $N_d/N$ in the case of a Kaiser(6) window and 75\% overlap is asymptotically 0.69, starting at 0.53 for very low values of $N$. The difference between these values is due to the last three terms which can be neglected for $N>10$. However, due to the assumption on the data, this estimate will overestimate  the number of independent averages in the case of non-white correlated data, and in particular when it is periodic.

A typical advice regarding the required number of independent averages is found in  \citet{carter1973estimation} where it is said that the statistical properties \textit{"... dramatically portray(s) the requirement that $N_d$ be large."} What this means in practice is that $N_d$ should preferrably be larger than 4-6 as in \citet{holm2014alleged} where the lowest value used was $N_d=7.1$. However, with a Monte Carlo based method for finding significance level, rather than an analytic one, lower values of $N_d$ can give meaningful results as demonstrated in this paper.


\newcommand{\noopsort}[1]{}

\end{document}